# Raman Study of Layered Breathing Kagome Lattice Semiconductor Nb$_3$Cl$_8$

Dylan A. Jeff[1,2], Favian Gonzalez[1,2], Kamal Harrison[1,2], Yuzhou Zhao[3,4], Tharindu Fernando[4], Sabin Regmi[1], Zhaoyu Liu[4], Humberto R. Gutierrez[5], Madhab Neupane[1], Jihui Yang[3], Jiun-Haw Chu[4], Xiaodong Xu[3,4], Ting Cao[3], Saiful I. Khondaker[1,2]*

[1]Department of Physics, University of Central Florida, Orlando, Florida 32816, United States
[2]NanoScience Technology Center, University of Central Florida, Orlando, FL 32826, USA
[3]Department of Materials Science and Engineering, University of Washington, Seattle, WA 98195, USA
[4]Department of Physics, University of Washington, Seattle, WA 98195, USA
[5]Department of Physics, University of South Florida, Tampa, FL 33620, USA
*Author to whom any correspondence should be addressed

E-mail: saiful@ucf.edu

**Abstract:**

Niobium chloride (Nb$_3$Cl$_8$) is a layered 2D semiconducting material with many exotic properties including a breathing kagome lattice, a topological flat band in its band structure, and a crystal structure that undergoes a structural and magnetic phase transition at temperatures below 90 K. Despite being a remarkable material with fascinating new physics, the understanding of its phonon properties is at its infancy. In this study, we investigate the phonon dynamics of Nb$_3$Cl$_8$ in bulk and few layer flakes using polarized Raman spectroscopy and density-functional theory (DFT) analysis to determine the material's vibrational modes, as well as their symmetrical representations and atomic displacements. We experimentally resolved 12 phonon modes, 5 of which are A$_{1g}$ modes while the remaining 7 are E$_g$ modes, which is in strong agreement with our DFT calculation. Layer-dependent results suggest that the Raman peak positions are mostly insensitive to changes in layer thickness, while peak intensity and FWHM are affected. Raman measurements as a function of excitation wavelength (473 – 785 nm) show a significant increase of the peak intensities when using a 473 nm excitation source, suggesting a near resonant condition. Temperature-dependent Raman experiments carried out above and below the transition temperature did not show any change in the symmetries of the phonon modes, suggesting that the structural phase transition is likely from the high temperature *P$\bar{3}$m*1 phase to the low-temperature *R$\bar{3}$m* phase. Magneto-Raman measurements carried out at 140 and 2 K between -2 to 2 T show that the Raman modes are not magnetically coupled. Overall, our study presented here significantly advances the fundamental understanding of layered Nb$_3$Cl$_8$ material which can be further exploited for future applications.

KEYWORDS: niobium halide, Raman spectroscopy, 2D materials, quantum material, kagome lattice

## 1. Introduction

Nb$_3$Cl$_8$ is a member of the niobium halide (Nb$_3X_8$, $X$ = Cl, Br, I) family of layered two-dimensional (2D) materials that crystallize in a breathing kagome crystal lattice [1]. They are shown to be semiconducting through angle-resolved photoemission spectroscopy (ARPES) measurements on bulk [2, 3], and have visible and infrared light absorbing capabilities. This family of materials recently garnered attention due to the effects that the breathing kagome lattice has on the material's band structure and magnetic orderings. Topological flat bands in the electronic band structures of $Nb_3Cl_8$ and $Nb_3I_8$ have been theoretically predicted and experimentally observed via ARPES [2, 3]. This exotic electronic band structure is linked to the unique geometry of the kagome crystal lattice, which causes the localization of electrons within the hexagons formed by the corners of the niobium trimers. One of the effects that the kagome lattice in this system has on the magnetic orderings takes the form of magnetic frustration, where the material does not feature a single lowest energy state for the orientation of its spins [4–6]. At low temperatures, the crystal symmetry of $Nb_3Cl_8$ changes due to the unequal distribution of electrical charge between different layers of Nb trimers in the lattice. This results in a buckling of Cl layers and a rearrangement of the stacking order, which overcomes the magnetic frustration. As a result, Nb3Cl8 forms a nonmagnetic singlet ground state. This change in crystal symmetry is seen as the crystal changes from trigonal *P$\bar{3}$m*1 at high temperatures to a predominantly monoclinic *C2/m* at low temperatures[4]. However, other studies suggested that the low-temperature phase is *R$\bar{3}$m* symmetric instead [5,7]. Additionally, $Nb_3Cl_8$ features favorable cleavage energies for the separation of layers from the bulk crystal, speaking to the feasibility of its exfoliation to thinner layers [8]. As a semiconductor with a moderate band gap of 1.12 eV, great light absorption, and increased conductivity in thin layers; $Nb_3Cl_8$ is a prime candidate for future electronic, optoelectronic, and spintronic device applications [2, 8]. Despite being an interesting material with many exotic electronic and magnetic properties, experimental investigations of the optical and vibrational properties of $Nb_3Cl_8$ such as the number of phonon modes, their symmetry assignments, atomic displacements, thickness dependence, and the effect of magnetic field and temperature on the phonon modes has not yet been studied. Such studies are significant in advancing the fundamental understanding of layered $Nb_3Cl_8$ and other members of the niobium halide family materials for the realization of many overarching goals in future applications.

Raman spectroscopy is used across several disciplines of science and engineering as a method of characterizing materials according to their molecular or lattice vibrations. Raman scattering results from the complex interaction between photons from an excitation source and the specific electrons and phonons in a material. Therefore a Raman spectrum is unique to each material, likening it to a "fingerprint" of the material [9–11]. Layer-dependent Raman studies on mechanically exfoliated layered materials are useful for finding contrasts between the crystal structure of a material's bulk and thin counterparts [12–17]. Vibrational symmetries can be studied through polarization-dependent Raman spectroscopy, providing insight into the symmetries of each Raman active phonon mode in the material. Different combinations of the polarization of the excitation laser and the scattered light can be matched to the angle-dependent intensities of the phonon modes to determine their symmetries [18, 19], which then enables one to probe for structural transitions by comparing the symmetries of the phonon modes before and after the transition. The manipulation of conditions such as temperature and external magnetic field is also a reliable method of further classifying features in a characteristic Raman spectrum and determining their origins. For example, in magnetic 2D materials such as $CrI_3$, $FePS_3$, $GdTe_3$, and $BiSe_2/Te_3$, both temperature dependent and magneto-Raman spectroscopy have been utilized to study the effects of temperature-dependent magnetic transitions and the possible contributions of magnons in the material's Raman signature [20–23]. Varying excitation wavelength can induce second order Raman scattering via a double resonance process where the energy of the incident light matches certain electronic transitions in the material, and the momentum of the involved phonons allows quasi-horizontal electronic transitions between electronic states in different points of the Brillouin zone [24]. By exploiting resonance conditions, one can resolve Raman modes that would otherwise have intensities too low to be resolved in non-resonant conditions [25–27].

Here, we investigate the phonon dynamics of Nb$_3$Cl$_8$ in bulk and few layer flakes using Raman spectroscopy and density-functional theory (DFT) analysis to determine the material's vibrational modes, their symmetrical representation, and their atomic displacements. The manipulation of parameters such as the polarization and excitation energy of the incident beam, temperature, and external magnetic field when performing the Raman measurement provided the means to study the response of the material's phonon modes to these changes and determine the nature of the Raman active modes. We experimentally resolved 12 phonon modes, 5 of which are A$_{1g}$ symmetric and 7 of which are E$_g$ symmetric. The calculated frequencies and symmetrical representations from our first principles DFT analysis agreed with our experimental observations. From the DFT results, we determined the atomic displacements of the atoms in the crystal lattice due to the Raman scattering event. Layer-dependent studies show that the Raman peak positions are mostly insensitive to changes in layer number, only slightly red shifting while peak intensity decreased and the full width at half maximum (FWHM) increased with layer thickness. Raman measurements at different excitations (473 – 785 nm) show the largest peak intensities when using a 473 nm excitation source, suggesting resonance between the laser's energy and the energy of an electronic transition in the material. Temperature dependent Raman measurements carried out above and below transition temperature using a 532 nm laser did not show any changes in the symmetries of the phonon modes when compared to the high temperature phase, suggesting that the likely low-temperature phase is $R\bar{3}m$ symmetric. Magneto-Raman measurements carried out at 140 and 2K between -2 to 2 T reveal that Raman modes are not magnetically coupled due to the absence of peak splitting, shift, or intensity loss.

## 2. Experimental section
### 2.1 Crystal growth and characterization

Single crystals of Nb$_3$Cl$_8$ were synthesized via chemical vapor transport [6] (adapted to two-zone furnace) using NH$_4$Cl as a transport agent. 0.1114 g Nb, 0.3639 g NbCl$_5$, and 0.0059 g NH$_4$Cl powders were sealed in a fused silica tube (12.75 mm OD × 10.5 mm ID) as precursors under vacuum and placed into a two-zone furnace. The furnace was ramped to 200°C for 60 minutes with a dwelling time of 200 minutes. The precursor side and growth side were then ramped to 830°C and 785°C respectively for 600 min with a dwelling time of 8600 min. The precursor and growth side were then cooled to 20°C and 350°C respectively for 420 min and dwelled for 180 min before cooling both sides to 20°C for 60 min. The grown Nb$_3$Cl$_8$ crystals were dark, flat, and shiny with a noticeable hexagonal shape.

Single crystal x-ray diffraction (SCXRD) measurements were carried out using a high-resolution diffractometer (Bruker D8 Discover) with Cu Kα radiation (λ = 1.5418 Å). The samples were placed on a zero-background silicon sample holder. Data were collected in the 2θ range of 10° to 90° with a step size of 0.02° and a counting time of 10 seconds per step.

Magnetic susceptibility (χ) was measured using the vibrating sample magnetometer (VSM) option on a physical property measurement system (PPMS) (Quantum Design PPMS DynaCool). Samples were weighed and mounted on a non-magnetic quartz paddle sample holder using GE-7031 varnish. The same holder with only varnish was used as a blank to subtract backgrounds. Measurements were performed at temperatures ranging from 2 K to 300 K with applied magnetic fields. At each temperature, the magnetic moment was measured as a function of magnetic field by sweeping the field from -5 to 5 T, and the slope between -0.5 to 0.5 T was fitted to extract the magnetic susceptibility of the sample at the temperature.

**2.2 Mechanical exfoliation of $Nb_3Cl_8$**

$Nb_3Cl_8$ flakes were mechanically exfoliated from the grown bulk crystals directly onto silicon wafers with a 250nm $SiO_2$ capping layer using Nitto tape (SPV-224 PVC). The flakes to be studied were then identified via optical microscopy, where color contrast and transparency were key factors in choosing the thin-layer flakes. Atomic force microscopy (AFM) was utilized to measure the thickness and identify the layer number of each flake studied during the experiment. AFM measurements taken in non-contact mode on a SmartSPM1000 scanning probe microscope. AFM measurements were taken in ambient conditions.

**2.3 Raman characterization**

The layer-dependent and polarized Raman experiments were performed on a confocal microscope Raman spectrometer (LabRAM HR Evolution, Horiba Scientific) using a backscattering geometry. The Raman emission was collected and dispersed by a 1800 gr/mm grating, using a 532 nm wavelength laser as the excitation source. The laser was focused on the exfoliated $Nb_3Cl_8$ flakes through a 100x objective (NA = 0.9, WD = 0.21 mm) with a spot size of ~ 1 µm. For the polarized Raman experiments, the polarization of the incident laser was rotated using a combination of a fixed linear polarizer (horizontally) and a rotating half-wave plate that determined the linear polarization angle, the scattered Raman signal passed through a second polarizer (analyzer) that was kept horizontal. Excitation dependent Raman spectroscopy was performed on both bulk and thin layer flakes utilizing 473 nm, 532 nm, 633 nm, and 785 nm excitation sources to probe for a resonance Raman effect in the material. These Raman measurements were performed at room temperature under ambient conditions. For the analysis of the Raman spectra, Gaussian-Lorentzian fittings were used to fit the Raman peaks and deduce the peak FWHMs, frequencies, and intensities published in this work.

Low-temperature Raman spectroscopy was performed using a backscattering geometry in a closed-cycle helium cryostat (Montana C2), and the measurements were performed in the temperature range of 7.6 K to 300 K. Light from a 532-nm continuous laser (Torus 532, Laser Quantum) was focused by a 40x objective lens to a spot size of ~ 1 µm. A laser power of 1 mW was used, and the collected signal was dispersed by a spectrometer (Princeton Instruments HRS-500-MS) using a 1200 $mm^{-1}$ groove-density grating and detected by a liquid nitrogen-cooled charge-coupled device (CCD, Princeton Instruments PyloN:400BR_eXcelon) with an integration time of 10 seconds. BragGrate$^{TM}$ notch filters were used to filter out Rayleigh scattering.

Magneto-Raman measurements were carried out in a closed-cycle helium cryostat down to 2 K with split-coil conical magnet (Quantum Design Opticool) using an Andor SR-500i spectrometer and an Andor Newton DU920P BR CCD.

**2.4 Calculation method**

First-principles calculations were performed, utilizing the density functional theory approach as implemented in the Quantum ESPRESSO package [28]. We used norm-conserving scalar relativistic Optimized Norm-Conserving Vanderbilt Pseudopotential (ONCVPSP) [29, 30] and Perdew-Burke-Ernzerhof-type exchange-correlational functional [31], and the van der Waals correction included in the D2 formalism [32]. A primitive unit cell with two layers was fully optimized to get a = 6.75 Å and c = 12.43 Å. The convergence threshold on forces for ionic relaxation was $1\times10^{-4}$ eV/Å. A Hubbard potential of U = 1 eV was used on Nb d orbitals to account for the on-site interaction [33]. Calculations were done on a 10x10x3 Monkhorst-Pack k-point grid.

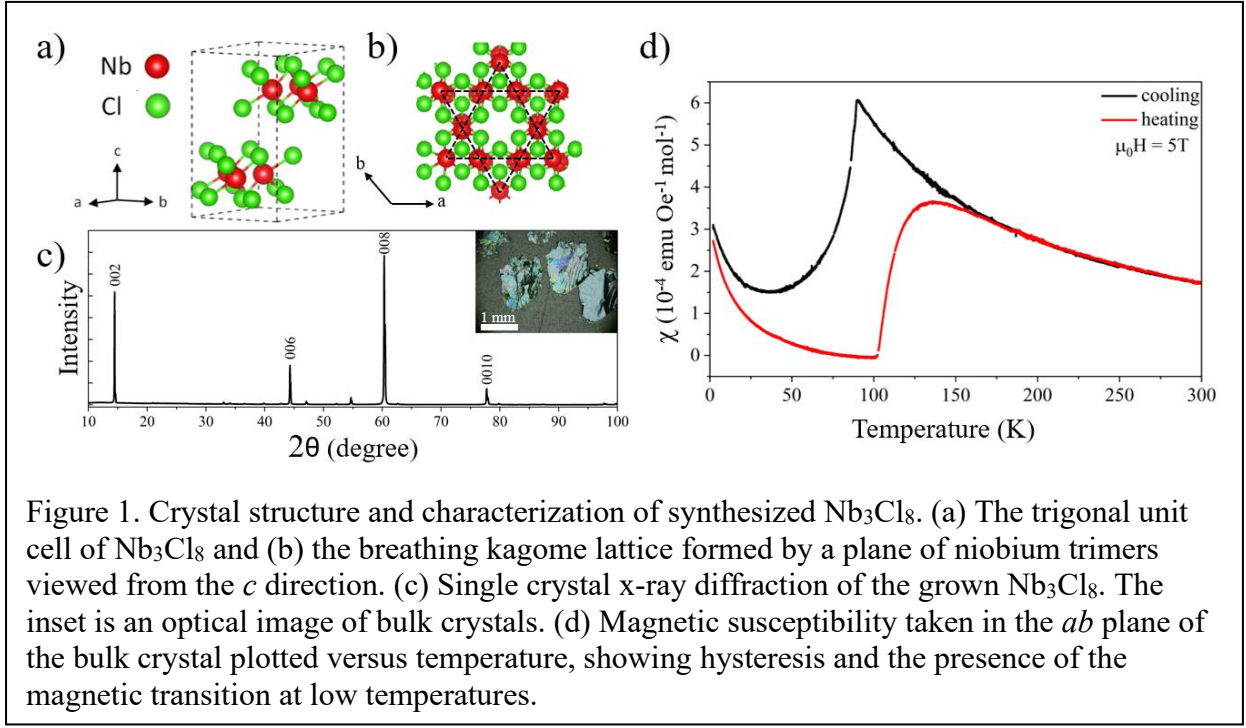

Figure 1. Crystal structure and characterization of synthesized $Nb_3Cl_8$. (a) The trigonal unit cell of $Nb_3Cl_8$ and (b) the breathing kagome lattice formed by a plane of niobium trimers viewed from the *c* direction. (c) Single crystal x-ray diffraction of the grown $Nb_3Cl_8$. The inset is an optical image of bulk crystals. (d) Magnetic susceptibility taken in the *ab* plane of the bulk crystal plotted versus temperature, showing hysteresis and the presence of the magnetic transition at low temperatures.

A density functional perturbation theory (DFPT) calculation was performed to calculate phonon modes. We note that electronic calculations were carried out for the interlayer antiferromagnetic stable ground state of the bulk material. However, for the Raman experiment at room temperature, the material does not have long-range magnetic order. As a result, we use nonmagnetic calculations to perform phonon calculations for the crystal system at 300 K. To test how spin polarizations would affect phonon frequencies, we performed additional calculations based on the interlayer antiferromagnetic ground states. Our calculations show that the shift of phonon energy is small (~0.2 meV on average).

## 3. Results and Discussion

The crystal structure of $Nb_3Cl_8$ as well as basic structural and magnetic characterizations of the bulk samples are summarized in Figure 1. $Nb_3Cl_8$ crystallizes in a trigonal crystal system with space group $P\bar{3}m1$ and point group $D_{3d}$. $Nb_3Cl_8$ has been reported to have a monolayer height of 0.8 nm on $SiO_2$/Si, with a vertical direction (out-of-plane) lattice constant of 0.61 nm in bulk [2]. As shown in Figure 1b, the $Nb_3Cl_8$ crystal features a breathing Kagome lattice formed by corner sharing Niobium trimers with alternating Nb-Nb distances [34, 35]. Figure 1c showcases the results of the SCXRD measurements taken on the bulk $Nb_3Cl_8$, grown using the chemical vapor transport technique described in the methods section. The sharp peaks and minimal noise indicate the good crystallinity of the sample. The periodicity of the lattice in the vertical direction is calculated to be 0.61 nm, consistent with the reported values [2]. The magnetic susceptibility of the grown crystal was then probed in a temperature cycle where the sample was first cooled down to 8 K and subsequently heated back to room temperature within a vibrating sample magnetometer (VSM). The magnetic susceptibility of the material with respect to temperature in Figure 1d shows hysteresis between the cooling and heating transitions, with the critical cooling temperature found to be at 90 K and the critical heating temperature found to be at 103K, both corresponding to the transition from a paramagnetic state at room temperature to a nonmagnetic singlet state [5,6] at low temperatures. This magnetic transition has been linked to the structural phase transition which $Nb_3Cl_8$ undergoes at low-temperatures in the literature, and therefore the presence of this magnetic transition in our data is the evidence that the structural transition has occurred [4–7].

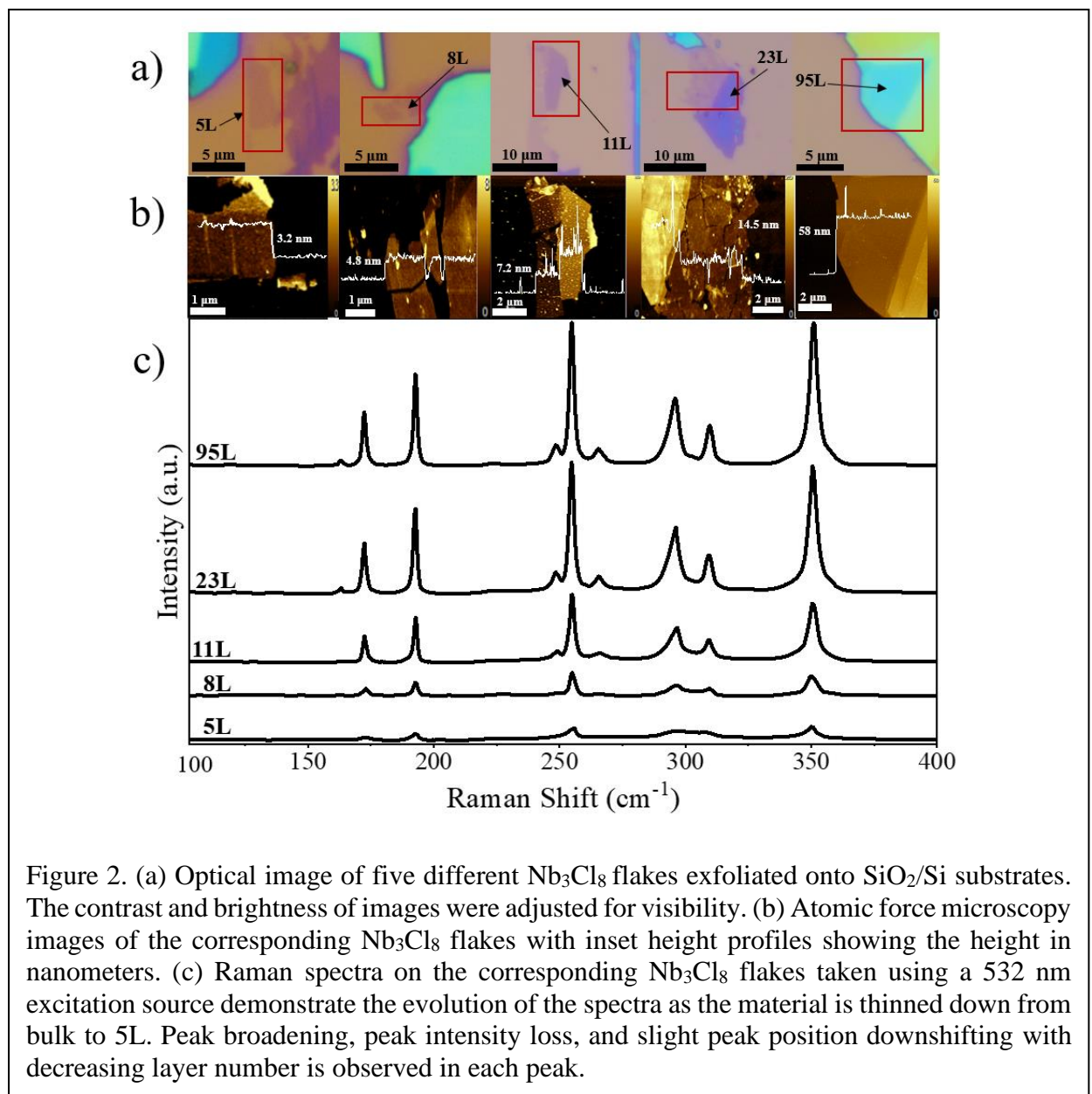

Figure 2. (a) Optical image of five different $Nb_3Cl_8$ flakes exfoliated onto $SiO_2$/Si substrates. The contrast and brightness of images were adjusted for visibility. (b) Atomic force microscopy images of the corresponding $Nb_3Cl_8$ flakes with inset height profiles showing the height in nanometers. (c) Raman spectra on the corresponding $Nb_3Cl_8$ flakes taken using a 532 nm excitation source demonstrate the evolution of the spectra as the material is thinned down from bulk to 5L. Peak broadening, peak intensity loss, and slight peak position downshifting with decreasing layer number is observed in each peak.

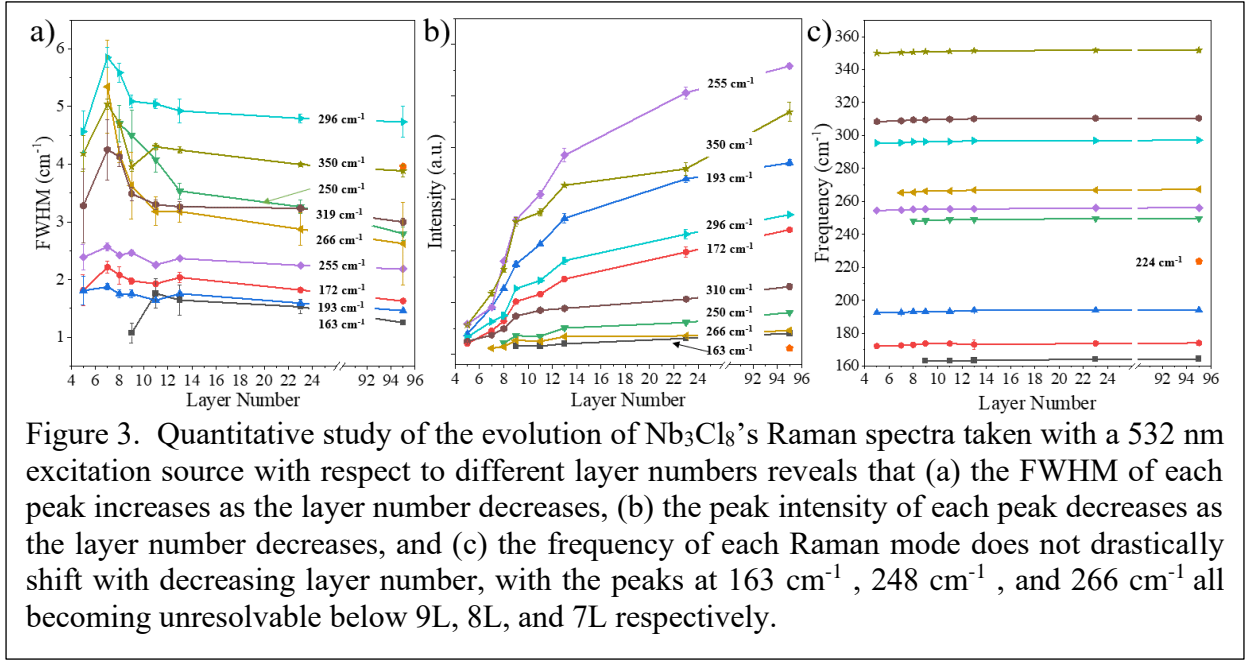

Figure 3. Quantitative study of the evolution of $Nb_3Cl_8$'s Raman spectra taken with a 532 nm excitation source with respect to different layer numbers reveals that (a) the FWHM of each peak increases as the layer number decreases, (b) the peak intensity of each peak decreases as the layer number decreases, and (c) the frequency of each Raman mode does not drastically shift with decreasing layer number, with the peaks at 163 cm$^{-1}$, 248 cm$^{-1}$, and 266 cm$^{-1}$ all becoming unresolvable below 9L, 8L, and 7L respectively.

### 3.1 Layer-dependent Raman analysis

$Nb_3Cl_8$ was mechanically exfoliated down to few layers using the techniques outlined in the methods section. Optical images, heights measured by atomic force microscopy (AFM), and Raman spectra using a 532 nm excitation source for 5 different $Nb_3Cl_8$ flakes with various layer numbers are shown in Figure 2. Figure 2a features the optical micrographs of flakes exfoliated onto a silicon substrate, showing that the material is a light blue color at 95 layers (95L), dark purple at 23L, and a light transparent purple as it is thinned down to 5L. Using the previously cited heights, on $SiO_2$ and in bulk, the AFM images in Figure 2b are used to assign layer numbers to each exfoliated flake. Changes in the Raman spectra with respect to layer number are evident in Figure 2c, where all peaks lose intensity and feature an increase in FWHM as the material is thinned down from bulk to thin layers. The Raman modes at 163 cm$^{-1}$, 250 cm$^{-1}$, and 266 cm$^{-1}$ all become unresolvable in flakes thinner than 7L when using a 532 nm excitation source at room temperature.

A quantitative study of $Nb_3Cl_8$'s Raman signature at room temperature using a 532 nm excitation source and its evolution as the material is thinned down from bulk is shown in Figure 3. As evident from Figure 3a, the FWHM of each peak generally increases by a minimum of about 1 cm$^{-1}$ as the layer number decreases, with a sharp decline in 5L due to the low intensity of each peak which is just above the noise level. Figure 3b exemplifies how the intensity of each Raman peak is closely related to the layer number of the flake upon which the spectrum was gathered, with the intensity decreasing as the material is thinned. Figure 3c reveals that none of the peaks experienced a drastic shift in wavenumber when studied from bulk to thin layers of $Nb_3Cl_8$, but all peaks slightly downshifted as layer number decreased (see supporting information Fig. S2). These redshifts can be explained by interlayer Van der Waals interactions which contribute less to the phonon restoring forces as the layer number decreases, resulting in softened phonon modes [15, 37]. Figure 3c also shows that the peaks at 163 cm$^{-1}$, 250 cm$^{-1}$, and 266 cm$^{-1}$ are not resolved in samples thinner than 7L, while the peak at 224 cm$^{-1}$ is unresolvable in thin layers. The disappearance of the peaks at 163 cm$^{-1}$, 224 cm$^{-1}$, 250 cm$^{-1}$, and 266 cm$^{-1}$ as well as the decrease in intensity with layer number can be linked to the effect that an increased surface-to-volume ratio (and effectively smaller volume) will have on the cross section of evoking a Raman event. The lower probabilities to induce the Raman events will result in the already less intense Raman modes being drowned out by noise as the material is thinned.

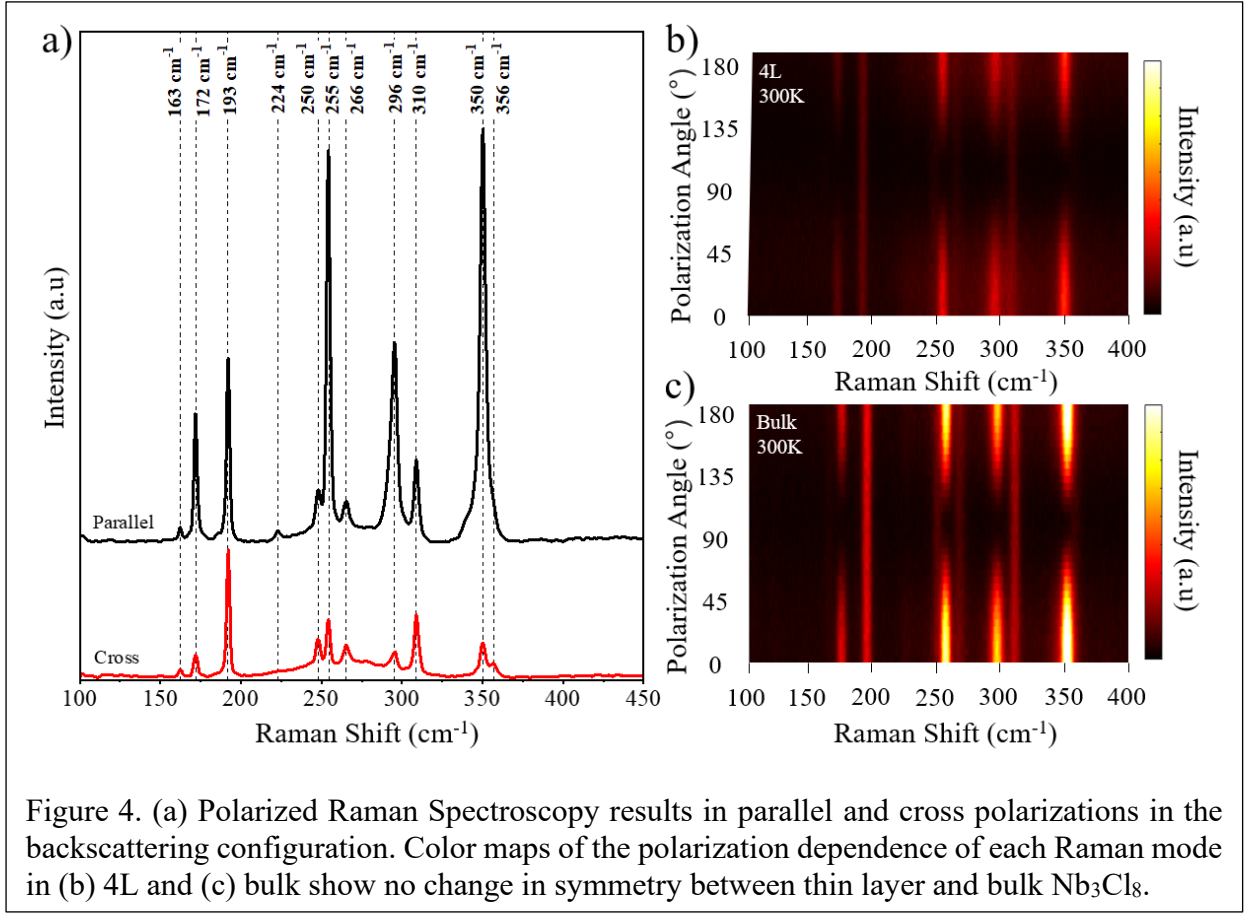

Figure 4. (a) Polarized Raman Spectroscopy results in parallel and cross polarizations in the backscattering configuration. Color maps of the polarization dependence of each Raman mode in (b) 4L and (c) bulk show no change in symmetry between thin layer and bulk Nb$_3$Cl$_8$.

### 3.2 Room and low-temperature polarized Raman analysis

The Raman spectra of bulk Nb$_3$Cl$_8$ measured using a 532 nm excitation source at room temperature in both parallel and cross polarizations in a backscattering configuration are shown in Figure 4. As a crystal with space group $P\bar{3}m1$, Nb$_3$Cl$_8$ is expected to have 19 Raman active phonon modes [37–39]. The irreducible representations of the Raman active optical phonons in bulk Nb$_3$Cl$_8$ at the Γ point of the Brillouin zone can be expressed as $\Gamma = 8A_{1g} + 11E_g$. Of these 19 expected phonon modes, 11 phonon modes were experimentally resolved using a 532 nm laser (Figure 4a) while an additional mode at 120 cm$^{-1}$ was observed using a 473 nm laser (see Figure 7), making for a total of 12 experimental phonons. While the results of both polarization conditions only show 10 different peaks, the cross polarized configuration (red curve in Figure 4a) showcases a shoulder peak at 356 cm$^{-1}$ that is revealed due to the intensity loss of the 350 cm$^{-1}$ peak, revealing the 11th phonon mode when using a 532 nm laser. Figures 4b and 4c provide a full Raman peak intensity map at each wavenumber as the polarization angle of the light is changed from 0 to 180 degrees to determine the symmetry of each phonon mode in 4L and bulk $Nb_3Cl_8$, respectively. These results show no difference in the phonon symmetries between the 4L and bulk sample, indicating a robust crystal phase regardless of layer number. As a member of the $P\bar{3}m1$ crystal space group, the Raman polarization selection rules for $Nb_3Cl_8$'s corresponding point group, $D_{3d}$, were used to assign phonon peaks to their respective symmetrical representations [37–39]. The $D_{3d}$ polarization selection rules indicate that the Raman spectra of $Nb_3Cl_8$ consist of $A_{1g}$ and $E_g$ symmetric modes. The $A_{1g}$ modes are "allowed" when the light reaching the Raman analyzer is linearly polarized parallel to the excitation source and "forbidden" upon cross polarization [18, 19, 41], while the $E_g$ modes are allowed for both configurations. Therefore, in the case where the only governing symmetries of the material's vibrational modes are $A_{1g}$ and $E_g$, a comparison between the Raman spectra produced by linearly polarized and cross polarized light is sufficient to determine the symmetries of the modes present in the Raman spectra. Due to drastic changes in the relative peak intensity upon swapping polarization configurations, the peaks at 172 $cm^{-1}$, 224 $cm^{-1}$, 255 $cm^{-1}$, 296 $cm^{-1}$, and 350 $cm^{-1}$ were determined to be representative of $A_{1g}$ symmetrical vibrations. While the peaks at 163 $cm^{-1}$, 193 $cm^{-1}$, 250 $cm^{-1}$, 266 $cm^{-1}$, 310 $cm^{-1}$, and 356 $cm^{-1}$ were found to be representative of $E_g$ vibrations due to their consistent intensity across all polarization configurations. These behaviors are seen in Figures 4b and 4c, most notably when tracking each peak's intensity at the 90-degree polarization angle, where $E_g$ peaks have consistent intensities while $A_{1g}$ peaks do not.

In the interest of probing the expected structural transition that $Nb_3Cl_8$ undergoes at temperatures lower than 90 K, polarized Raman spectroscopy above and below the transition temperature utilizing a 532 nm excitation source was performed on bulk and thin layer samples. Figure 5 details the phonon dynamics of the crystal system at low-temperatures (7.6K) with respect to changes in polarization. In comparison to the high-temperature data from Figure 4, the Raman peak intensities with respect to the polarization angles do not change in the low-temperature data, indicating that the symmetries of the phonons do not change as the crystal undergoes the structural phase transition. To interpret these results, we must revisit the controversy, mentioned earlier, that exists in the literature regarding the low-temperature phase. The structural transition was first observed in literature via diffraction data, showing that the material transitions from crystallizing at high temperature in the $P\bar{3}m1$ space group (point group $D_{3d}$) to crystallizing at low temperature in the predominantly $C2/m$ space group (point group $C_{2h}$)[4]. In this case, there is a significant change in the type of Raman active phonons expected for the low-temperature phase ($A_g$ and $B_g$) as well as the Raman tensors governing the polarization selection rules[37,38,39]. However, other groups have reported that the low-temperature phase corresponds to the $R\bar{3}m$ space group instead [5, 7]. Since crystals with $P\bar{3}m1$ and $R\bar{3}m$ symmetries share the same point group ($D_{3d}$) and hence the same Raman tensors for the $A_g$ and $E_g$ modes, there would be no difference in the polarization-dependence of these modes. The lack of evidence of new phonons with $B_g$ symmetry in the 7.6 K data shown in Figure 5, as well as the unchanged polarization-dependence of the Raman modes is more consistent with the low-temperature phase being $R\bar{3}m$ rather than $C2/m$.

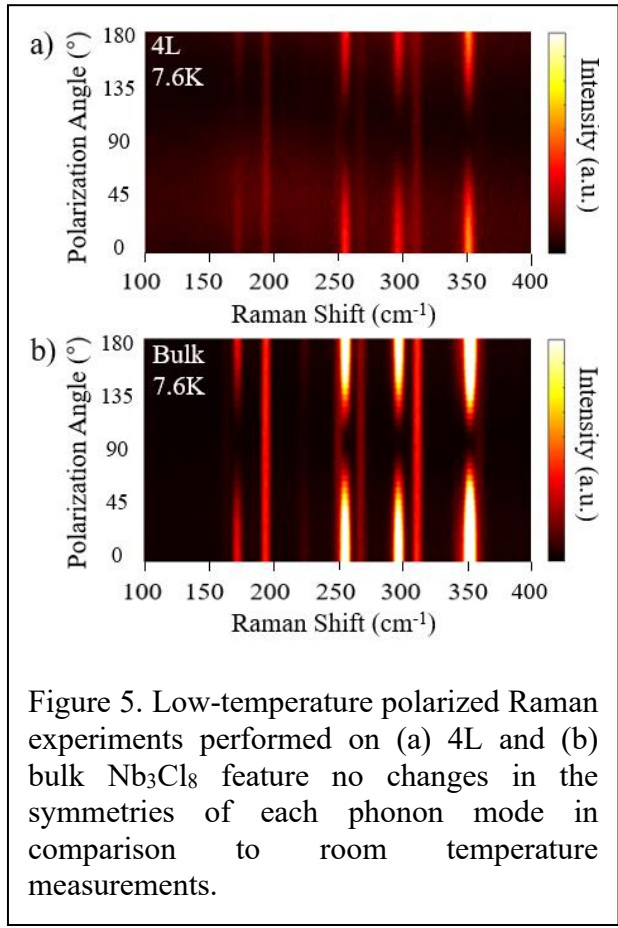

Figure 5. Low-temperature polarized Raman experiments performed on (a) 4L and (b) bulk $Nb_3Cl_8$ feature no changes in the symmetries of each phonon mode in comparison to room temperature measurements.

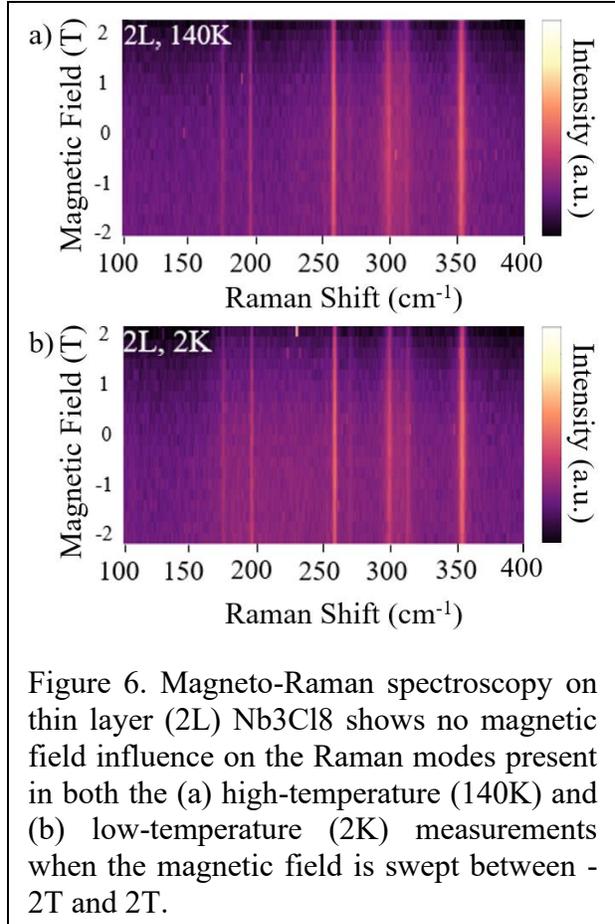

Figure 6. Magneto-Raman spectroscopy on thin layer (2L) Nb3Cl8 shows no magnetic field influence on the Raman modes present in both the (a) high-temperature (140K) and (b) low-temperature (2K) measurements when the magnetic field is swept between -2T and 2T.

A recent terahertz study on $Nb_3Cl_8$ showed that an IR-active mode at 104 cm$^{-1}$ experiences an abrupt shift in frequency to 108 cm$^{-1}$ as the system crossed the transition temperature of 90K, which was claimed to be evidence of the structural phase transition from $P\bar{3}m1$ to $R\bar{3}m$ phase. We have also observed a shift in peak frequency of up to 5 cm$^{-1}$ between 300 K and 7.6 K data for our measured peaks between 163 cm$^{-1}$ to 350 cm$^{-1}$ measured using a 532 nm laser (see supporting information Fig. S4). To investigate if the shift in frequency is abrupt as observed in ref [7], we present additional Raman peak position data at 140 K and 80 K for a 2L sample (see supporting information Fig. S5). We note that the experiment was conducted using a different setup (Opticool) with distinct lasers and a spectrometer compared to the Raman experiment carried out at 300K and 7.6K (Montana C2). Nevertheless, under the same experimental conditions, we observed a maximum shift of about 1 cm$^{-1}$ when comparing the 140K data to the 80K data. Although, it is not clear if a significant shift in the peak frequencies is expected between the high temperature $P\bar{3}m1$ and low temperature $R\bar{3}m$ phase for the frequency range we investigated, we cannot rule out the possibility of peak shifts in the Raman modes due to the structural phase transition. We also believe that thermal contraction of the crystal structure as the temperature decreased, has an effect on the Raman frequency shifts observed between 300 and 7.6 K.

### 3.3 Magneto-Raman analysis

To study the influence that an external magnetic field could have on the observed Raman modes of $Nb_3Cl_8$, magneto-Raman spectroscopy was carried out on a thin (2L) $Nb_3Cl_8$ flake, both above (140 K) and below (2 K) the magnetic and structural phase transition temperature of 90 K. This measurement can elucidate the presence of magnon peaks, as it has been shown in literature that materials with magnetic orderings in thin layers should feature drastic peak splitting, shifting, or intensity loss [20, 21, 42, 43]. As the external magnetic field was swept between -2 T and 2 T, the intensities and frequencies of each peak were tracked. As seen in the contour plots in Figure 6 and supplementary figure S6 in the supporting information, there was no change in the intensities or frequencies of any of the Raman modes as the magnetic field strength was swept. Furthermore, no difference is seen between the high-temperature measurement in Figure 6a and the low temperature measurement in Figure 6b, revealing that the magnetic transition also has no effect on the peak positions or general behaviors of Nb$_3$Cl$_8$'s Raman modes in the presence of an external magnetic field.

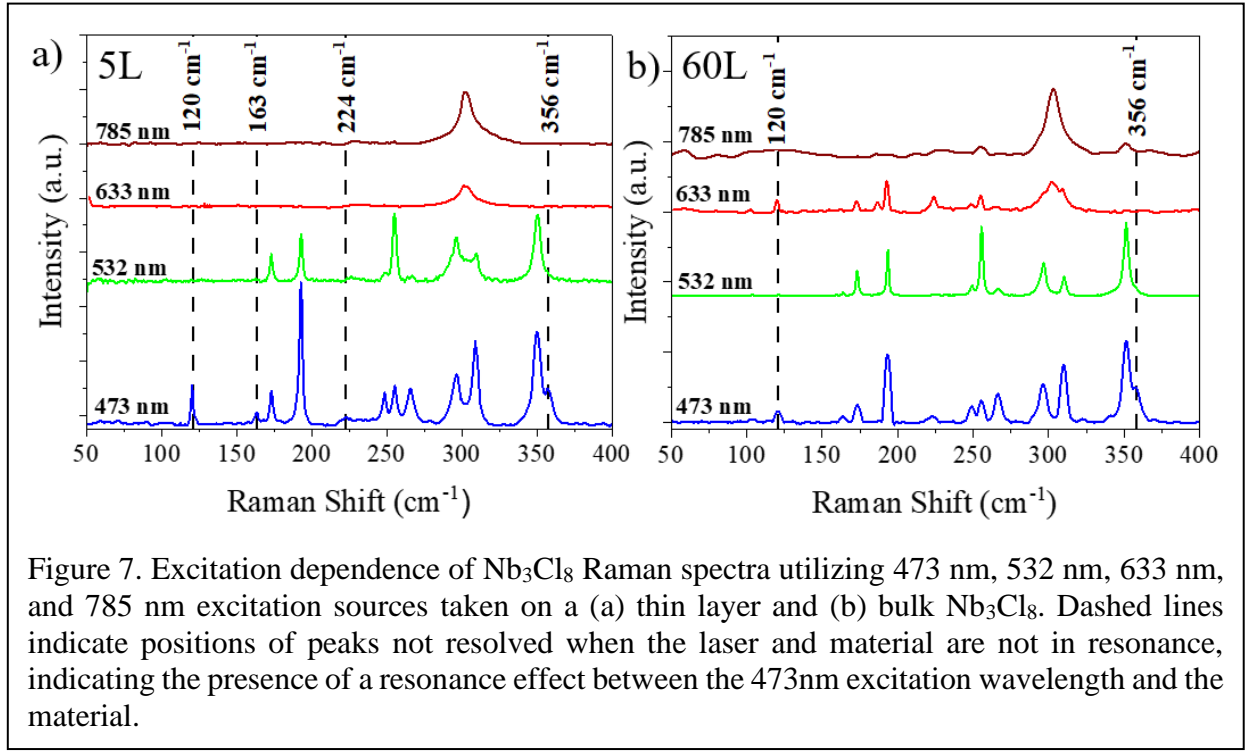

Figure 7. Excitation dependence of Nb$_3$Cl$_8$ Raman spectra utilizing 473 nm, 532 nm, 633 nm, and 785 nm excitation sources taken on a (a) thin layer and (b) bulk Nb$_3$Cl$_8$. Dashed lines indicate positions of peaks not resolved when the laser and material are not in resonance, indicating the presence of a resonance effect between the 473nm excitation wavelength and the material.

### 3.4 Excitation-dependent Raman analysis

With absorption peaks in the visible light range, a resonant Raman effect can be expected in Nb$_3$Cl$_8$ when probed with excitation sources in the visible range. Particularly, the absorption spectrum of Nb$_3$Cl$_8$ features a broad peak between 500 nm and 400 nm [1, 2]. Considering this, an excitation-dependent Raman study in the parallel polarization configuration on the material using lasers with wavelengths in the visible range was performed in pursuit of the resonant Raman effect. The results of the excitation-dependent Raman experiment on thin layer (5L) and bulk (60L) Nb$_3$Cl$_8$ are presented in Figures 7a and 7b, respectively. Raman spectra were taken on the bulk and thin layer samples with four different excitation sources to probe for a resonance effect between the energies of the pumped lasers and electronic transitions within the material. In comparison to the parallel polarized spectra taken with a 532 nm source from Figure 4, the 473 nm laser resolved peaks at 120 cm$^{-1}$, 163 cm$^{-1}$, 224 cm$^{-1}$, and 356 cm$^{-1}$ even in thin layers. This increase in peak intensity in thin layers is evidence of the resonance effect, allowing us to resolve some of the least intense phonon modes in $Nb_3Cl_8$' Raman signature. Meanwhile, the Raman spectra produced by the 633 and 785 nm lasers are more difficult to resolve phonons in both bulk and thin layers. This can be due to large optical reflectance peaks $Nb_3Cl_8$ exhibits in the 600 to 800 nm range (shown in supporting information Fig. S3.).

### 3.5 Density functional theory analysis

Table 1: Calculated and experimental Raman active phonon modes of $Nb_3Cl_8$

| Calculated | | Experimental | |
|---|---|---|---|
| Frequency (cm$^{-1}$) | Irreducible Representation | Frequency (cm$^{-1}$) | Irreducible Representation |
| 27.1 | $E_g$ | | |
| 48.9 | $A_{1g}$ | | |
| 102.2 | $E_g$ | | |
| 122.3 | $E_g$ | 120.4 | |
| 164.8 | $E_g$ | 163.7 | $E_g$ |
| 177.6 | $A_{1g}$ | 173.6 | $A_{1g}$ |
| 189.4 | $E_g$ | 193.2 | $E_g$ |
| 193.5 | $A_{1g}$ | | |
| 216.2 | $A_{1g}$ | 224.5 | $A_{1g}$ |
| 216.9 | $E_g$ | | |
| 237.4 | $E_g$ | 250.1 | $E_g$ |
| 252.6 | $A_{1g}$ | 255.4 | $A_{1g}$ |
| 258.4 | $E_g$ | 266.3 | $E_g$ |
| 280.5 | $E_g$ | | |
| 285.4 | $A_{1g}$ | 296.3 | $A_{1g}$ |
| 301.8 | $E_g$ | 310.4 | $E_g$ |
| 339.1 | $A_{1g}$ | 350.2 | $A_{1g}$ |
| 348.3 | $E_g$ | 356.8 | $E_g$ |
| 396.1 | $A_{1g}$ | | |

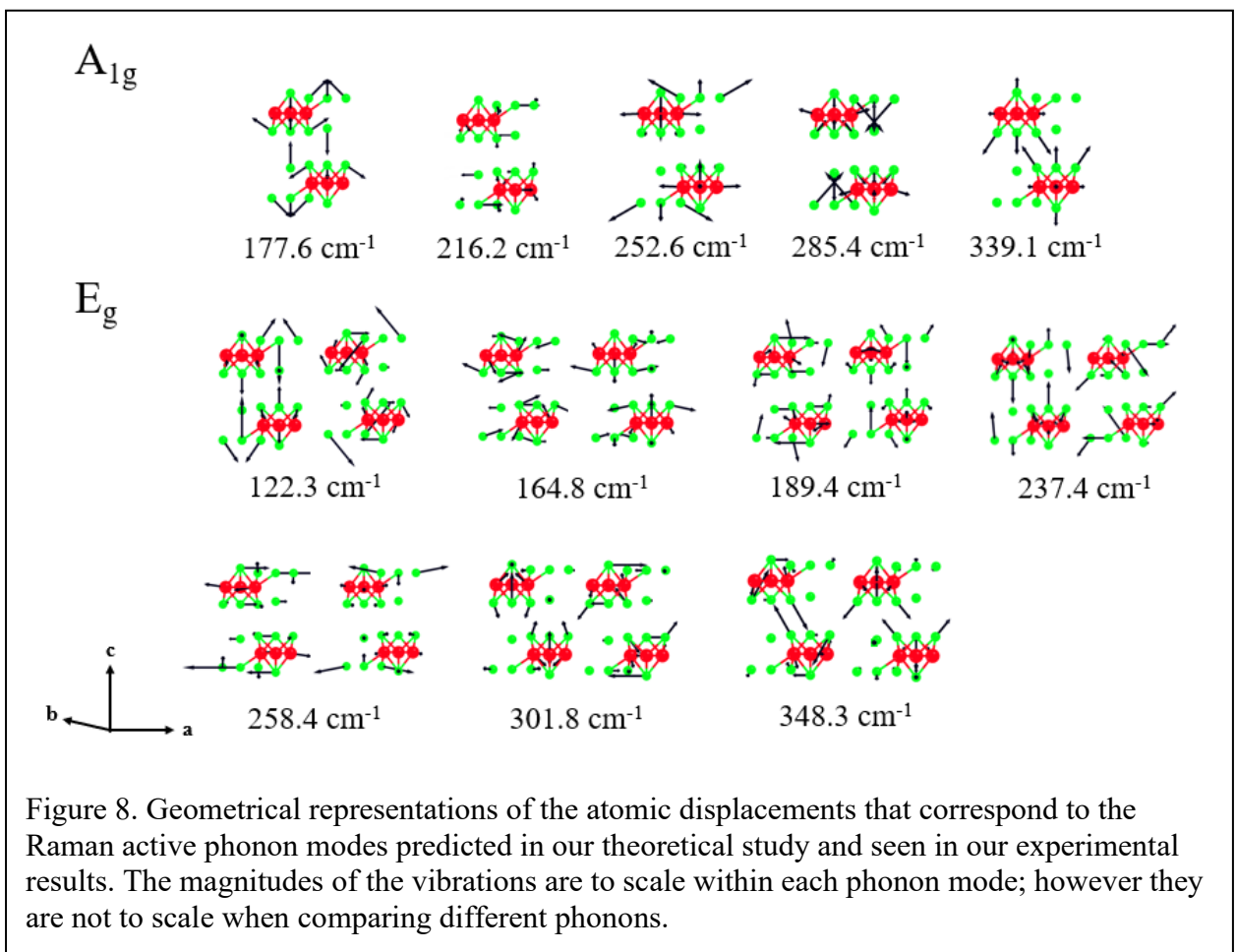

Figure 8. Geometrical representations of the atomic displacements that correspond to the Raman active phonon modes predicted in our theoretical study and seen in our experimental results. The magnitudes of the vibrations are to scale within each phonon mode; however they are not to scale when comparing different phonons.

     First-principles DFT calculations were utilized to compute the symmetries and frequencies of the 19 Raman active phonon modes in $Nb_3Cl_8$ at 300 K. Our results revealed that 8 of the modes exhibit $A_g$ symmetry while the other 11 exhibit $E_g$ symmetry, which is in accordance with the expected number of optical phonons determined via our Raman tensor analysis. Table 1 summarizes our experimental and theoretical findings for each phonon mode, providing a comparison between the observed experimental frequencies and symmetries and those predicted via our DFT calculations. We found that the calculated values are in reasonable agreement with the experimental values from our Raman studies. The DFT analysis further allowed us to assign a vibrational symmetry of $E_g$ to the 120 cm$^{-1}$ mode, which was only experimentally resolved under resonance conditions (see Fig. 7). Figure 8 provides a visualization of $Nb_3Cl_8$'s experimentally resolved phonon modes and their corresponding atomic displacements. We modeled the atomic displacements of the 12 phonon modes and found that the 7 $E_g$ modes are vibrationally degenerate, while the other 5 are non-degenerate $A_{1g}$ modes. We accentuate this distinction by showing the $A_{1g}$ and $E_g$ modes separately. Notably, the degeneracy is a result of two distinct yet orthogonal vibrational modes that correspond to a single phonon. This orthogonality between the vibrations of the degenerate phonons clarifies the polarization angle-independent intensities of the $E_g$ modes. The contrast between non-degenerate $A_{1g}$ and degenerate $E_g$ modes is further exemplified in our

## 4. Conclusion

To summarize, the effects of layer thickness, polarization, temperature, magnetic field, and excitation energy on the phonon dynamics of $Nb_3Cl_8$ have been investigated. Utilizing Raman spectroscopy, 12 phonon modes were resolved, of those 12, 5 were determined to be $A_{1g}$ symmetric and 7 were determined to be $E_g$ symmetric. The frequencies, symmetrical assignments, and atomic displacements of the phonon modes calculated at room temperature via our first principles DFT analysis agreed with these experimental results. The layer-dependent studies revealed that the phonon frequencies only slightly decreased with layer number, while intensity decreased and FWHM increased more notably with layer number. Polarized Raman experiments at low-temperature show that the phonon modes do not experience any changes in symmetry when compared to those of the high-temperature phase, suggesting that the structural phase transition is likely from high temperature $P\bar{3}m1$ phase to low-temperature $R\bar{3}m$ phase. Magneto Raman measurements show the phonon peaks did not experience any shift indicating the magnetic transition has no effect on $Nb_3Cl_8$'s Raman modes in the presence of an external magnetic field. The Raman measurements taken at various excitation wavelengths revealed that the 473 nm excitation source exhibited the greatest resonance effect out of all the wavelengths tested. Our results provide a foundational understanding of the phonon dynamics of $Nb_3Cl_8$ under a variety of conditions, giving insight into the fundamental optical and vibrational properties of the crystal. as well as elucidating the effects that the low temperature structural transition has on the material's vibrational symmetries.

**Acknowledgments**

This work was supported by Grant DMR-2121953 from NSF Partnerships for Research and Education in Materials (PREM).

We acknowledge the support from UW Molecular Engineering Materials Center, an NSF Materials Research Science and Engineering Center (Grant No. DMR-1719797).

The DFT analysis was facilitated through the use of advanced computational, storage, and networking infrastructure provided by the Hyak supercomputer system and funded by the University of Washington Molecular Engineering Materials Center at the University of Washington (DMR-1719797).

We further acknowledge the materials project database [43] and the visualization for electronic and structural analysis (VESTA) software [44] for providing the data and models necessary for the portrayal of the crystal structure's unit cell utilized in this work.